\documentclass[aps,prc,showpacs,showkeys,superscriptaddress,eqsecnum]{revtex4}

\usepackage{graphicx}
\usepackage{amsmath}
\usepackage{amsfonts}
\usepackage{amssymb}
\usepackage{bm}

\newcommand{\beq}{\begin{equation}}
\newcommand{\eeq}{\end{equation}}
\newcommand{\beqa}{\begin{eqnarray}}
\newcommand{\eeqa}{\end{eqnarray}}

\newcommand{\nn}{\nonumber \\ }

\date{\today}
\begin{document}

\title{Dominant contributions to the nucleon-nucleon interaction at sixth 
order of chiral perturbation theory}

\author{D. R. Entem}
\email{entem@usal.es}
\affiliation{Grupo de F\'isica Nuclear, IUFFyM, Universidad de Salamanca, 
E-37008 Salamanca, Spain}
\author{N. Kaiser}
\email{nkaiser@ph.tum.de}
\affiliation{Physik Department T39 , Technische Universit\H{a}t M\H{u}nchen, 
D-85747 Garching, Germany}
\author{R. Machleidt}
\email{machleid@uidaho.edu}
\affiliation{Department of Physics, University of Idaho, Moscow, Idaho 83844, 
USA}
\author{Y. Nosyk}
\affiliation{Department of Physics, University of Idaho, Moscow, Idaho 83844, 
USA}

\begin{abstract}
We present the dominant two- and three-pion-exchange contributions to the 
nucleon-nucleon interaction at sixth order 
(next-to-next-to-next-to-next-to-next-to-leading order, N$^5$LO) of chiral 
perturbation theory. Phase shifts with orbital angular momentum $L\geq4$ are 
given parameter free at this order and allow for a systematic investigation 
of the convergence of the chiral expansion. The N$^5$LO contribution is 
prevailingly repulsive and considerably smaller than the N$^4$LO one, thus, 
establishing the desired trend towards convergence. Using low-energy 
constants that were extracted from an analysis of $\pi N$-scattering at fourth 
order, the predictions at N$^5$LO are in excellent agreement with the 
empirical phase shifts of peripheral partial waves.
\end{abstract}

\maketitle

\section{Introduction}
\label{sec_intro}

The derivation of nuclear forces from chiral effective field theory has been 
a topic of active research for the past quarter century~\cite{Wei90,ORK94,
KBW97,KGW98,Kai00a,Kai00b,Kai01,Kai01a,Kai02,EM02,EM03,EGM98,EGM05,Eks13,
Ent15,Pia15,Kai15} (see also Refs.~\cite{ME11,EHM09} for recent reviews). 
By 1998, the evaluation of the nucleon-nucleon ($N\!N$) interaction up to 
next-to-next-to-leading order (N$^2$LO, third order in small momenta) was 
completed~\cite{ORK94,KBW97,KGW98} and, by 2003, these calculations were  
extended to N$^3$LO~\cite{Kai00a,Kai00b,Kai01,Kai01a,Kai02,EM02,EM03}. As it 
turned out, at N$^2$LO and N$^3$LO, one is faced with a surplus of attraction, 
in particular, when the low-energy constants (LECs) for subleading 
pion-nucleon couplings are applied consistently as extracted from analyses 
of elastic $\pi N$-scattering~\cite{KBW97,KGW98,EM02}. Finally, in 2014, this 
issue was picked up and calculations up to N$^4$LO were conducted~\cite{Ent15}.
It was shown that the $2\pi$- and $3\pi$-exchange contributions at N$^4$LO 
are prevailingly repulsive and, thus, are able to fully compensate the 
excessive attraction of the lower orders. However, it was also noticed that 
the N$^2$LO, N$^3$LO, and N$^4$LO contributions are all roughly of the same 
magnitude, raising legitimate concerns about the convergence of the chiral
expansion of the $N\!N$-potential. 

It is, therefore, the purpose of the present paper to move on to the next 
order and to investigate the $N\!N$-interaction at N$^5$LO (of sixth power in 
small momenta) with the goal to obtain more insight into the convergence 
issue. 

Besides this, the order N$^5$LO has other interesting features. At this order, 
a new set of $N\!N$-contact terms depending with the sixth power on momenta 
appears, bringing the total number of short-distance parameters to 50. 
This set includes then terms that contribute up to $F$-waves.

However, the focus of the present paper is on peripheral partial waves with
orbital angular momentum $L\geq 4$, which are exclusively ruled by the 
non-polynomial pion-exchange expressions constrained by chiral symmetry.
Hence, this investigation is a test of the implications of chiral symmetry 
for the $N\!N$-interaction up to sixth order.

This paper is organized as follows:
In Secs.~IIA, IIB, and IIC, we consider the two-, three-, and four-pion 
exchange contributions at sixth order and argue that some parts are 
negligibly small. The predictions for elastic $N\!N$-scattering in peripheral 
partial waves are shown in Sec.~III, and Sec.~IV concludes the paper. 

\section{Pion-exchange contributions to the $N\!N$-interaction 
at N$^5$LO}\label{sec_pions}

This section is subdivided into three subsections in which we will consider 
various classes of two- and three-pion exchange diagrams. We will present 
arguments for neglecting the chiral four-pion exchange at this order. Our 
semi-analytical results will be stated in terms of contributions to the 
momentum-space $N\!N$-amplitudes in the center-of-mass system (CMS), 
which arise from the following general decomposition of the $N\!N$-potential:
\begin{eqnarray} 
V({\vec p}~', \vec p\,) &  = &
 \:\, V_C \:\, + \bm{\tau}_1 \cdot \bm{\tau}_2 \, W_C 
\nonumber \\ & & + 
\left[ \, V_S \:\, + \bm{\tau}_1 \cdot \bm{\tau}_2 \, W_S 
\,\:\, \right] \,
\vec\sigma_1 \cdot \vec \sigma_2
\nonumber \\ &&+ 
\left[ \, V_{LS} + \bm{\tau}_1 \cdot \bm{\tau}_2 \, W_{LS}    
\right] \,
i \vec S \cdot (\vec k \times \vec q\,)
\nonumber \\ &&+ 
\left[ \, V_T \:\,     + \bm{\tau}_1 \cdot \bm{\tau}_2 \, W_T 
\,\:\, \right] \,
\vec \sigma_1 \cdot \vec q \,\, \vec \sigma_2 \cdot \vec q  
\nonumber \\ &&+ 
\left[ \, V_{\sigma L} + \bm{\tau}_1 \cdot \bm{\tau}_2 \, 
      W_{\sigma L} \, \right] \,
\vec\sigma_1\cdot(\vec q\times \vec k\,) \,\,
\vec \sigma_2 \cdot(\vec q\times \vec k\,)
\, ,
\label{eq_nnamp}
\end{eqnarray}
where ${\vec p}\,'$ and $\vec p$ denote the final and initial nucleon momenta 
in the CMS, respectively. Moreover, $\vec q = {\vec p}\,' - \vec p$ is the 
momentum transfer, $\vec k =({\vec p}\,' + \vec p)/2$ the average momentum, 
and $\vec S =(\vec\sigma_1+\vec\sigma_2)/2 $ the total spin, with 
$\vec \sigma_{1,2}$ and $\bm{\tau}_{1,2}$ the spin and isospin operators, of 
nucleon 1 and 2, respectively. For on-shell scattering, $V_\alpha$ and 
$W_\alpha$ ($\alpha=C,S,LS,T,\sigma L$) can be expressed as functions of 
$q= |\vec q\,|$ and $k=|{\vec k}|$, only. The one-pion exchange contribution 
is of the well-known form $W_T^{(1\pi)} = - (g_A/2f_\pi)^2 (m_\pi^2+q^2)^{-1}$ with 
with $g_A$ the axial-vector coupling constant, $f_\pi=92.4\,$MeV the pion 
decay constant, and $m_\pi$ the pion mass. Numerical values for $g_A$ and 
$m_\pi$ will be given in Sec.~III. This expression fixes at the same time 
our sign-convention for the $N\!N$-potential $V({\vec p}~', \vec p)$.

We will state contributions in terms of their spectral functions, from which
the momentum-space amplitudes $V_\alpha(q)$ and $W_\alpha(q)$ are obtained
via the subtracted dispersion integrals:
\begin{eqnarray} 
V_{C,S}(q) &=&  {2 q^8 \over \pi} \int_{nm_\pi}^{\tilde{\Lambda}} d\mu \,
{{\rm Im\,}V_{C,S}(i \mu) \over \mu^7 (\mu^2+q^2) }\,, 
\nn
V_T(q) &=&  -{2 q^6 \over \pi} \int_{nm_\pi}^{\tilde{\Lambda}} d\mu \,
{{\rm Im\,}V_T(i \mu) \over \mu^5 (\mu^2+q^2) }\,, 
\label{eq_disp}
\end{eqnarray}
and similarly for $W_{C,S,T}$. Clearly, the thresholds are given by $n=2$ for 
two-pion exchange and $n=3$ for three-pion exchange. For $\tilde{\Lambda} 
\rightarrow \infty$ the above dispersion integrals yield the finite parts of 
loop-functions as in dimensional regularization, while for finite 
$\tilde{\Lambda} >> nm_\pi$ we employ the method known as spectral-function 
regularization (SFR) \cite{EGM04}. The purpose of the finite scale 
$\tilde{\Lambda}$ is to constrain the imaginary parts to the  
low-momentum region where chiral effective field theory is applicable.

\subsection{Two-pion exchange contributions at N$^5$LO}
\label{sec_2pi}

\begin{figure}
\scalebox{0.8}{\includegraphics{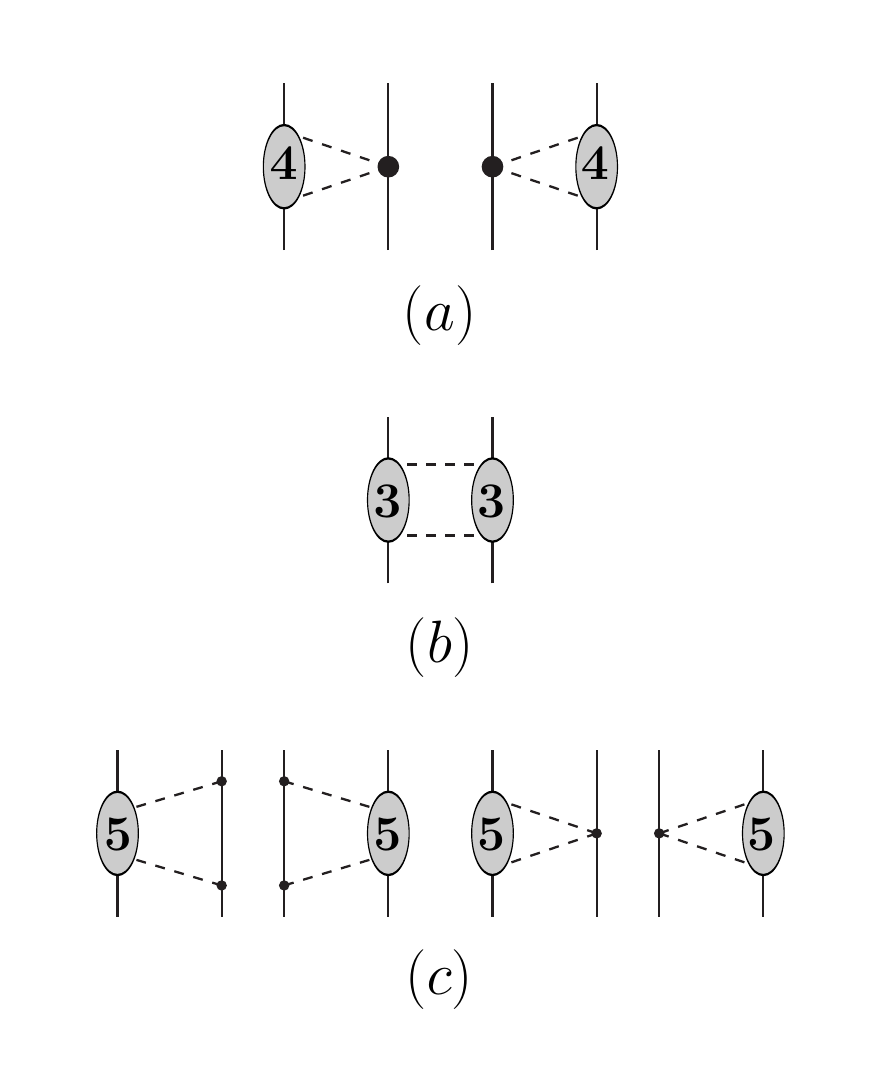}}
\vspace*{-0.5cm}
\caption{Two-pion-exchange contributions to the $N\!N$-interaction at N$^5$LO.
(a) The subleading one-loop $\pi N$-amplitude is folded with the chiral 
$\pi\pi NN$-vertices proportional to $c_i$. (b) The leading one-loop 
$\pi N$-amplitude is folded with itself. (c) The leading two-loop 
$\pi N$-amplitude is folded with the tree-level $\pi N$-amplitude. Solid lines
represent nucleons and dashed lines pions. Small dots and large solid dots
denote vertices of chiral order one and two, respectively. Shaded ovals 
represent complete $\pi N$-scattering amplitudes with their order 
specified by the number in the oval.}
\label{fig_dia1}
\end{figure}

The $2\pi$-exchange contributions that occur at N$^5$LO are displayed 
graphically in Fig.~\ref{fig_dia1}. We will now discuss each class separately.

\subsubsection{Spectral functions for $2\pi$-exchange class (a)}
\label{sec_2pia}

The N$^5$LO $2\pi$-exchange two-loop contributions, denoted by class (a),  
are shown in Fig.~\ref{fig_dia1}(a). For this class the spectral functions 
are obtained by integrating the product of the subleading one-loop 
$\pi N$-amplitude (see Ref.~\cite{KGE12} for details) and the chiral 
$\pi\pi NN$-vertex proportional to $c_i$ over the Lorentz-invariant 
$2\pi$-phase space. In the $\pi\pi$ center-of-mass frame this integral can 
be expressed as an angular integral $\int_{-1}^1 dx$ \cite{Kai01a}. Altogether,
the results for the non-vanishing spectral functions read:
\begin{eqnarray} 
{\rm Im}V_C &\!\!\!\!=\!\!\!\!&{m_\pi^6 \sqrt{u^2-4}\over
(8\pi f_\pi^2)^3}\bigg({1\over
u^2}-2\bigg)\Big[(c_2+6c_3)u^2+4(6c_1-c_2-3c_3)
\Big] \Bigg\{2c_1 u+{c_2 u\over 36}(5u^2-24)\nonumber \\ &&+{c_3 u\over 2}
(u^2-2)+\bigg[c_3(2-u^2)+{c_2\over 6}(4-u^2)-4c_1\bigg]\sqrt{u^2-4}\,B(u)
\Bigg\} \nonumber \\ && + {m_\pi^6 \sqrt{u^2-4} \over 8 \pi
f_\pi ^4 u} \Bigg\{\Big[4c_1+c_3(u^2-2)\Big]\bigg[\bar e_{15}
(u^4-6u^2+8)+6\bar e_{14}(u^2-2)^2+{3\bar e_{16} \over 10}(u^2-4)^2\bigg]
\nonumber \\ && +c_2(u^2-4)\bigg[{3\bar e_{15} \over 10}(u^4-6u^2+8)
+\bar e_{14}(u^2-2)^2+{3\bar e_{16} \over 28}(u^2-4)^2\bigg] \Bigg\}\,,
\label{eq_2pia1}
\end{eqnarray}

\begin{eqnarray}
{\rm Im}W_S &\!\!\!\!=\!\!\!\!&{c_4^2 m_\pi^6 (u^2-4)\over 9(8\pi f_\pi^2)^3} 
\Bigg\{u\sqrt{u^2-4}\bigg[{5u^2 \over 6}-4+{2g_A^2 \over 15}(2u^2-23)\bigg]
-(u^2-4)^2 B(u)\nonumber \\ && +6g_A^2u \int_0^1\!dx\,\bigg(x-{1\over
x}\bigg) \Big[4+(u^2-4)x^2\Big]^{3/2} \ln{x \sqrt{u^2-4}+\sqrt{4+(u^2-4)x^2}
\over 2}\Bigg\}\nonumber \\ &&+{c_4 m_\pi^6 u(u^2-4)^{3/2} \over 240 \pi f_\pi^4}
\Big[10 \bar e_{17}(2-u^2)+\bar e_{18}(4-u^2)\Big]= \mu^2\, {\rm Im}W_T\,,
\label{eq_2pia2}
\end{eqnarray} 
with the dimensionless variable $u=\mu/m_\pi > 2$ and the logarithmic function
\begin{equation}
B(u) = \ln{u+\sqrt{u^2-4}\over 2}\,.
\end{equation}

Consistent with the calculation of the $\pi N$-amplitude in Ref.~\cite{KGE12}, 
we utilized the relations between the fourth-order LECs, such that only 
$\bar e_{14}$ to $\bar e_{18}$ remain in the final result.

\subsubsection{Spectral functions for $2\pi$-exchange class (b)}
\label{sec_2pib}
A first set of $2\pi$-exchange contributions at three-loop order, denoted by 
class (b), is displayed in Fig.~\ref{fig_dia1}(b). For this class of 
diagrams, the leading one-loop $\pi N$-scattering amplitude is multiplied 
with itself and integrated over the $2\pi$-phase space. Including also the 
symmetry factor $1/2$, one gets for the spectral-functions: 

\begin{eqnarray} 
{\rm Im}V_C &\!\!\!\!=\!\!\!\!&{m_\pi^6 \sqrt{u^2-4}\over
(4f_\pi)^8\pi^3u}\Bigg\{-{3\over 70}(5u^2+8)(u^2-4)^2+3g_A^2(1-2u^2)\bigg[
1+{2-u^2\over 4u} \ln{u+2\over u-2}\bigg]  \nonumber \\ && \times \bigg[
u-{u^3\over 2} +{4B(u) \over \sqrt{u^2-4}}\bigg] +g_A^4\bigg[{32(3-2u^2)
\over \sqrt{u^2-4}}B(u)+3(2u^2-1)^2\bigg({u^2-2 \over u}\ln{u+2 \over u-2}
\nonumber \\ &&+{(u^2-2)^2\over 8u^2} \bigg(\pi^2-\ln^2{u+2\over u-2}
\bigg)\bigg)-{2258\over 35}+24u+{5336u^2\over 105}-12u^3-{2216u^4\over 105}
+{18u^6 \over 35} \bigg]\nonumber \\ && +g_A^6(2u^2-1)\bigg(1+{2-u^2
\over 4u}
\ln{u+2\over u-2} \bigg) \bigg[46u-3u^3-96+{64\over u+2}+{24(5-2u^2)\over
\sqrt{u^2-4}}B(u)\bigg]\nonumber \\ && +{64g_A^8 \over
9}\bigg[{3119u^2\over
70}-{71u^6 \over 1120}-{197u^4\over 70}-{85u^3\over 8}+{97u\over 4}
-{582\over 7}-{16\over u+2} +{8\over (u+2)^2}\nonumber \\ &&
+{6u^4-60u^2+105\over \sqrt{u^2-4}}B(u)\bigg]\Bigg\}\,,
\end{eqnarray}

\begin{eqnarray} 
{\rm Im}W_S
&\!\!\!\!=\!\!\!\!&{g_A^4m_\pi^6\sqrt{u^2-4}\over
(4f_\pi)^8\pi^3 u} \Bigg\{{u^2-4\over 48}\bigg[4u+(4-u^2)\ln{u+2 \over u-2}
\bigg]^2-{\pi^2\over 48}(u^2-4)^3 \nonumber \\ && +g_A^2u \bigg[(u^2-4)
\ln{u+2 \over u-2}-4u\bigg]\bigg[{5u\over 4}-{u^3\over 24}-{8\over 3}
+{5-u^2\over \sqrt{u^2-4}}B(u) \bigg] \nonumber \\ && +{32g_A^4u^2\over 27}
\bigg[{u^4\over 40}+{13u^2\over 10}+{11u\over 2}-{118 \over 5}
-{8\over u+2}+{3(10-u^2)\over \sqrt{u^2-4}}B(u)\bigg]\Bigg\}
= \mu^2 {\rm Im}W_T\,, 
\end{eqnarray}

\begin{eqnarray} 
{\rm Im}V_S &\!\!\!\!=\!\!\!\!&{g_A^8m_\pi^6 u
\sqrt{u^2-4}\over
3(4f_\pi)^8\pi^5}\int_0^1\!dx\,(x^2-1)
\Bigg\{(u^2-4)x\bigg[{48\pi^2f_\pi^2\over
g_A^4}(\bar d_{14}-\bar d_{15})-{1\over 6}\bigg]+{4\over x}\nonumber \\ &&
  -{\big[4+(u^2-4)x^2\big]^{3/2}\over x^2\sqrt{u^2-4}} \ln{x
\sqrt{u^2-4}+\sqrt{
4+(u^2-4)x^2}\over 2}\Bigg\}^2 = \mu^2 {\rm Im}V_T\,,
\end{eqnarray}

\begin{eqnarray} 
{\rm Im}W_C &\!\!\!\!=\!\!\!\!&-{m_\pi^6(u^2-4)^{5/2}
\over
(4f_\pi)^8(3\pi u)^3}\bigg[2+4g_A^2-{u^2\over 2}(1+5g_A^2)\bigg]^2+ {m_\pi^6
(u^2-4)^{3/2}\over 9(4f_\pi)^8 \pi^5 u}
\int_0^1\!dx\,x^2\Bigg\{{3x^2\over 2}
(4-u^2) \nonumber \\ && +3x \sqrt{u^2-4}\sqrt{4+(u^2-4)x^2} \ln{x
\sqrt{u^2-4}
+\sqrt{4+(u^2-4)x^2}\over 2}+g_A^4\Big[(4-u^2)x^2\nonumber \\ &&+2u^2-4\Big]
\Bigg[{5\over 6}+{4\over (u^2-4)x^2}-\bigg(1+{4\over (u^2-4)x^2}
\bigg)^{3/2} \ln{x \sqrt{u^2-4}+\sqrt{4+(u^2-4)x^2}\over 2}
\Bigg]\nonumber \\ &&
+\Big[4(1+2g_A^2)-u^2(1+5g_A^2)\Big] \sqrt{u^2-4}\, {B(u) \over u}
+{u^2\over 6}
(5+13g_A^2)-4(1+2g_A^2) \nonumber \\ &&+96\pi^2f_\pi^2
\Big[(4-2u^2)(\bar d_1
+\bar d_2)+(4-u^2)x^2\bar d_3+8\bar d_5\Big] \Bigg\}^2\,.
\end{eqnarray}
Note the squared integrands in the last two equations. The parameters 
$\bar d_j$ belong to the $\pi\pi NN$-contact vertices of third chiral order.

\subsubsection{$2\pi$ class (c)}
\label{sec_2pic}

Further $2\pi$-exchange three-loop contributions at N$^5$LO, denoted by 
class (c), are shown in Fig.~\ref{fig_dia1}(c). For these the two-loop 
$\pi N$-scattering amplitude (which is of order five) would have to be folded 
with the tree-level $\pi N$-amplitude. To our knowledge, the two-loop elastic 
$\pi N$-scattering amplitude has never been evaluated in some decent 
analytical form. Note that the loops involved in the class (c) contributions 
include only leading order chiral $\pi N$-vertices. According to our 
experience such contributions are typically small. For these reasons
we omit class (c) in the present calculation.

\subsubsection{Relativistic $1/M_N^2$-corrections}
\label{sec_rel}

\begin{figure}
\scalebox{0.7}{\includegraphics{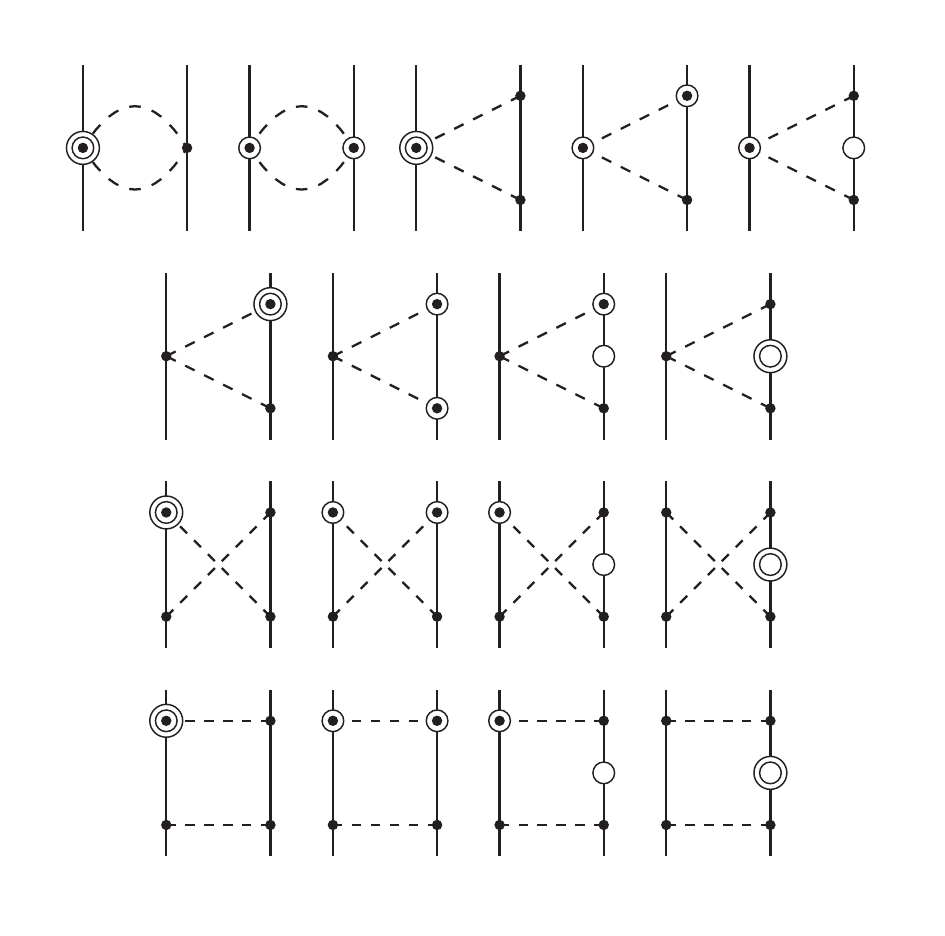}}
\caption{Relativistic $1/M_N^2$ corrections to $2\pi$-exchange diagrams that 
are counted as order six. Notation as in Fig.~\ref{fig_dia1}.
Open circles represent $1/M_N$-corrections.}
\label{fig_dia2}
\end{figure}

This group consists of the $1/M_N^2$-corrections to the chiral leading 
$2\pi$-exchange diagrams. Representative graphs are shown in 
Fig.~\ref{fig_dia2}. Since we count $Q/M_N \sim (Q/\Lambda_\chi)^2$, these 
relativistic corrections  are formally of sixth order (N$^5$LO). 
The expressions for the corresponding $N\!N$-amplitudes are adopted from 
Ref.~\cite{Kai02}:

\begin{eqnarray} 
V_C &=& {g_A^4 \over 32\pi^2 M_N^2 f_\pi^4}\bigg[ L(\tilde{\Lambda};q) \, 
\Big(2m_\pi^4+q^4 -8m_\pi^6 w^{-2}-2m_\pi^8 w^{-4}\Big)-{ m_\pi^6 \over 2 w^{2}}\, 
\bigg] ,
\label{eq_4M2C}
\\
W_C           &=& {1\over 192\pi^2 M_N^2 f_\pi^4} \bigg\{ L(\tilde{\Lambda};q) 
\bigg[g_A^2\Big(2k^2(8m_\pi^2+5q^2)+12m_\pi^6 w^{-2}-3 q^4-6m_\pi^2 q^2-6m_\pi^4
\Big) \nonumber \\   && + g_A^4 \Big(k^2(16m_\pi^4 w^{-2}-20m_\pi^2-7q^2) 
-16m_\pi^8 w^{-4}-12 m_\pi^6 w^{-2} +4m_\pi^4q^2w^{-2} +5q^4 +6m_\pi^2 q^2+6m_\pi^4 
\Big) \nonumber \\   && +k^2 w^2 \bigg]- {4 g_A^4 m_\pi^6 \over w^{2}} \bigg\} 
\,, \\
V_T &=& -{1\over q^2} V_S
    \; = \; {g_A^4 \, L(\tilde{\Lambda};q) \over 32\pi^2 M_N^2 f_\pi^4} 
        \bigg(k^2+{5\over 8} q^2 +m_\pi^4 w^{-2} \bigg) \,,
\\
W_T &=& -{1\over q^2} W_S 
    \; = \; { L(\tilde{\Lambda};q) \over 1536\pi^2 M_N^2 f_\pi^4}\bigg[
 g_A^4\Big(28m_\pi^2+17 q^2+16m_\pi^4 w^{-2} \Big)-2g_A^2(16 m_\pi^2+7q^2)
+ w^2 \bigg] \,, 
\nonumber \\
\\
V_{LS} &=& {g_A^4 \, L(\tilde{\Lambda};q) \over 128\pi^2 M_N^2 f_\pi^4} 
\Big( 11 q^2 +32m_\pi^4 w^{-2}\Big) \,,
\\
W_{LS} &=&  { L(\tilde{\Lambda};q) \over 256 \pi^2 M_N^2 f_\pi^4}
\bigg[2 g_A^2( 8m_\pi^2+3q^2)+ {g_A^4\over 3}\Big(16m_\pi^4 w^{-2}-11q^2 
-36m_\pi^2 \Big)-w^2 \bigg] \,,
\\
V_{\sigma L} &=& {g_A^4 \, L(\tilde{\Lambda};q) \over 32\pi^2 M_N^2 f_\pi^4}\;,
\label{eq_4M2sL}
\end{eqnarray} 
where the (regularized) logarithmic loop function is given by
\begin{equation} 
L(\tilde{\Lambda};q) = {w\over 2q} 
\ln {\frac{\tilde{\Lambda}^2(2m_\pi^2+q^2)-2m_\pi^2 q^2+\tilde{\Lambda}\sqrt{
\tilde{\Lambda}^2-4m_\pi^2}\, q\,w}{2m_\pi^2(\tilde{\Lambda}^2+q^2)}}\,,
\label{eq_L}
\end{equation}
with the abbreviation $ w = \sqrt{4m_\pi^2+q^2}$.

\subsection{Three-pion exchange contributions at N$^5$LO}
\label{sec_3pi}

\begin{figure}
\scalebox{0.7}{\includegraphics{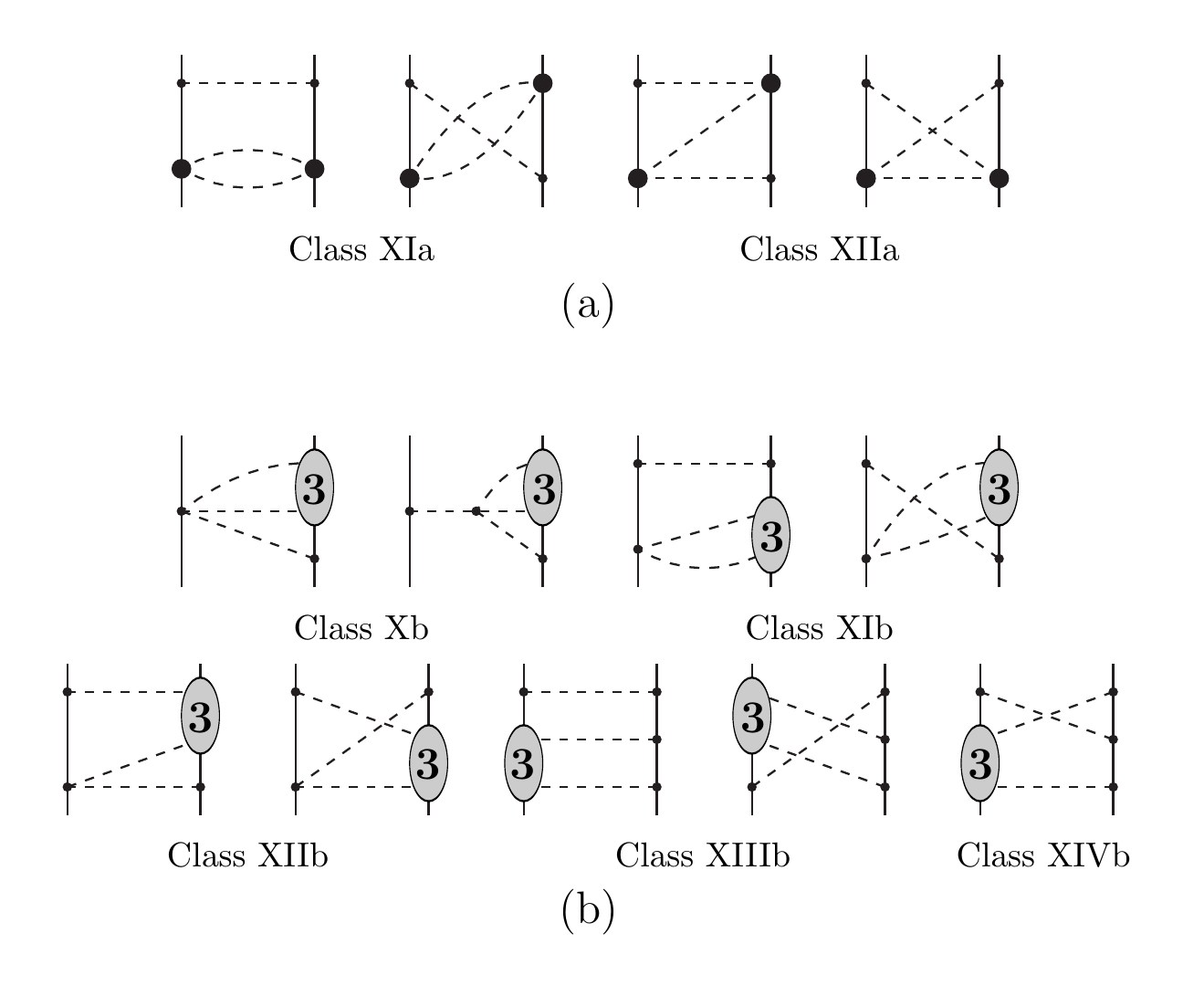}}
\caption{Three-pion exchange contributions at N$^5$LO.
(a) Diagrams proportional to $c_i^2$.
(b) Diagrams involving the one-loop $\pi N$-amplitude.
Roman numerals refer to sub-classes following the scheme introduced in  
Refs.~\cite{Kai01,Ent15}. Notation as in Fig.~\ref{fig_dia1}.}
\label{fig_dia3}
\end{figure}

The $3\pi$-exchange contributions of order N$^5$LO are shown in 
Fig.~\ref{fig_dia3}. We can distinguish between diagrams which are 
proportional to $c_i^2$ [Fig.~\ref{fig_dia3}(a)] and contributions that 
involve (parts of) the leading one-loop $\pi N$ amplitude 
[Fig.~\ref{fig_dia3}(b)]. Below, we present the spectral functions for each 
class.

\subsubsection{Spectral functions for $3\pi$-exchange class (a)}
\label{sec_3pia}

This class consists of the diagrams displayed in Fig.~\ref{fig_dia3}(a). They 
are characterized by the presence a subleading $\pi\pi NN$-vertices in each 
nucleon line. Using a notation introduced in Refs.~\cite{Kai01,Ent15},
we distinguish between the various sub-classes of diagrams by roman numerals.

Class XIa:

\begin{equation}{\rm Im}W_C = {g_A^2 c_4^2m_\pi^6\over  6(4\pi f_\pi^2)^3}
\int\limits_2^{u-1}\!dw\,(w^2-4)^{3/2}\sqrt{\lambda(w)}\,, \end{equation}

\begin{eqnarray} {\rm Im}V_S&\!\!\!\!=\!\!\!\!& {g_A^2c_4^2m_\pi^6\over
6(8\pi f_\pi^2)^3}\int\limits_2^{u-1}\!dw\,{(w^2-4)^{3/2}\over u^4
\sqrt{\lambda(w)}} \Big[w^8-4(1+u^2)w^6+2w^4(3+5u^2)\nonumber \\ &&
+4w^2(2u^6-5u^4-2u^2-1)-(u^2-1)^3(5u^2+1) \Big]\,, \end{eqnarray}

\begin{equation} {\rm Im}(\mu^2V_T-V_S)={g_A^2 c_4^2m_\pi^6\over 6(8\pi
f_\pi^2)^3} \int\limits_2^{u-1}\!dw\,(w^2-4)^{3/2}\sqrt{\lambda(w)}\,\bigg[
{(w^2-1)^2\over u^4}+1-{2\over u^2}(7w^2+1)\bigg]\,, \end{equation}
with the kinematical function $\lambda(w) = w^4+u^4+1-2(w^2u^2+w^2+u^2)$. The 
dimensionless integration variable $w$ is the invariant mass of a pion-pair 
divided by $m_\pi$. 

Class XIIa:

\begin{equation}{\rm Im}V_C = {g_A^2 c_4^2m_\pi^6\over 8960\pi f_\pi^6}\,
(u-3)^3\bigg[u^3+9u^2+12u-3- {3\over u}\bigg]\,, \end{equation}
\begin{equation}{\rm Im}W_C = {2g_A^2 c_4^2m_\pi^6u^2\over (4\pi f_\pi^2)^3}
\int\!\!\!\!\int\limits_{\!\!\!\!\!z^2<1}\!d\omega_1d\omega_2\, k_1 k_2
\sqrt{1-z^2} \arcsin(z)\,,\end{equation}

\begin{eqnarray} {\rm Im}V_S&\!\!\!\!=\!\!\!\!& {g_A^2 c_4^2m_\pi^6\over
(4\pi f_\pi^2)^3} \int\!\!\!\!\int\limits_{\!\!\!\!\!z^2<1}\!d\omega_1d\omega_2\,
\bigg\{2\omega_1^2(\omega_2^2-9\omega_2u+9u^2+1)+3\omega_1\big[\omega_2(1+8u^2)
-6u-6u^3\big]\nonumber \\ && +{1\over 4}(9u^4+18u^2+5)+{2z k_2\over k_1}\Big[
\omega_1^3(4u-\omega_2)+\omega_1^2(7\omega_2u-2-2u^2)-2\omega_1(2u+\omega_2)
\nonumber \\ && +2+2u^2-4\omega_2 u\Big]+{3 \arcsin(z)\over k_1k_2\sqrt{1-z^2}}
\bigg[2\omega_1^3u(u^2+1-2 \omega_2u)+\omega_1^2\Big(\omega_2u (7+11u^2) -
5\omega_2^2 u^2\nonumber \\ && -1-4u^2-3u^4\Big)+{\omega_1\over 4}
\Big(6u^5+12u^3-2u -\omega_2(5+16u^2+15u^4)\Big)+{(1-u^4)(u^2+3)\over 8}
\bigg]\bigg\}\,,\nonumber \\ &&  \end{eqnarray}

\begin{eqnarray} {\rm Im}(\mu^2V_T-V_S)&\!\!\!\!=\!\!\!\!& {g_A^2c_4^2m_\pi^6
\over (4\pi f_\pi^2)^3}\int\!\!\!\!\int\limits_{\!\!\!\!\!z^2<1}\!d\omega_1
d\omega_2\,\bigg\{ 4\omega_1^2(\omega_2^2+6u^2+2-10\omega_2 u)+6u^2(1+u^2) 
\nonumber \\ && +2\omega_1\big[3\omega_2(1+7u^2)-18u^3-10u\big] +{2z k_2 
\over k_1}\Big[ \omega_1^3(7u-2\omega_2)+u^2-\omega_2u\nonumber \\ &&
+\omega_1^2(13\omega_2 u -3-10u^2)+\omega_1(2+3u^2)(u-2\omega_2)\Big]
+{3 \arcsin(z)\over k_1k_2\sqrt{1-z^2}}\nonumber \\ && \times (u^2-2\omega_1u
+1)(u^2-2\omega_2u+1)\bigg[{\omega_1\over 2}(6u-5\omega_2) -{u^2\over 2}
-2\omega_1^2\bigg]\bigg\}\,,\end{eqnarray}

with the magnitudes of pion-momenta divided by $m_\pi$, and their 
scalar-product given by:
\begin{equation} k_1 = \sqrt{\omega_1^2 -1}\,, \qquad k_2 = \sqrt{\omega_2^2
-1}\,, \qquad z\, k_1k_2 = \omega_1\omega_2-u(\omega_1+\omega_2)+ {u^2+1
\over 2} \,.\end{equation}
The upper/lower limits of the $\omega_2$-integration are $\omega_2^\pm
= {1\over 2}(u-\omega_1\pm k_1 \sqrt{u^2-2\omega_1 u -3}/ \sqrt{u^2-2\omega_1
u +1}\,)$ with $\omega_1$ in the range $1<\omega_1<(u^2-3)/2u$.

The contributions to Im$W_S$ and Im$(\mu^2W_T-W_S)$ are split into three
pieces according to their dependence on the isoscalar/isovector low-energy
constants $c_{1,3}$ and $c_4$:

\begin{eqnarray} {\rm Im}W_S&\!\!\!\!=\!\!\!\!& {g_A^2m_\pi^6(u-3)^2
\over 2240 \pi f_\pi^6}\Bigg\{7c_1^2\bigg({4\over 3}+{3\over u}-{2\over
3u^2}-{1\over u^3} \bigg)+c_1 c_3\bigg({2u^2\over 3}+4u-{2\over 3}\nonumber 
\\ &&-{5\over u}-{2\over 3u^2}-{1\over u^3}\bigg)+c_3^2\bigg({3u^2\over 4}
+{u\over 8}-{5\over 2} -{3\over u}+{19\over 12u^2}+{19\over 8u^3}\bigg)
\Bigg\}\,,\end{eqnarray}

\begin{eqnarray} {\rm Im}(\mu^2W_T-W_S)&\!\!\!\!=\!\!\!\!& {g_A^2m_\pi^6(u-3)
\over 1120\pi f_\pi^6} \Bigg\{7c_1^2\bigg({1\over 3u}+{1\over u^2}+{3\over u^3}
-2u-1\bigg)+c_1 c_3\bigg(13u+4-5u^2-{5u^3\over 3} \nonumber \\ && +{1\over 3u}
+{1\over u^2}+{3\over u^3}\bigg)+{c_3^2\over 8}\bigg(23u^2-{u^5\over 3}-u^4-4u^3
-8u-3+{8\over 3u}-{19\over u^2}-{57\over u^3}\bigg)\Bigg\}\,,\nonumber \\ &&
\end{eqnarray}

\begin{eqnarray} {\rm Im}W_S&\!\!\!\!=\!\!\!\!& {g_A^2c_4 m_\pi^6 \over 1120\pi
f_\pi^6}(u-3)^2 \Bigg\{c_1\bigg(u^2+6u-1-{15\over 2u}-{1\over u^2}-{3\over 2u^3}
\bigg) \nonumber \\ && +{c_3\over 4}\bigg({2u^4\over 9}+{4u^3\over 3}+
{u^2\over 3}-{25u\over 6}+{6\over u}+{1\over u^2}+{3\over 2u^3}\bigg)
\Bigg\}\,,\end{eqnarray}

\begin{eqnarray} {\rm Im}(\mu^2W_T-W_S)&\!\!\!\!=\!\!\!\!& {g_A^2c_4m_\pi^6
\over 1120\pi f_\pi^6}(u-3)^3 \Bigg\{c_1\bigg({1\over u^2}+{1\over u^3}-
{u\over 3}-3-{4\over u}\bigg)  \nonumber \\ && +{c_3\over 4}\bigg({u^3\over 9}
+u^2+{5u\over 3}+{8\over 3}+{11\over 3u}-{1\over u^2}-{1\over u^3}
\bigg)\Bigg\}\,,\end{eqnarray}

\begin{equation} {\rm Im}W_S={g_A^2c_4^2m_\pi^6\over 8960\pi f_\pi^6}(u-3)^2
\bigg({25u\over 12}-{u^4\over 9}-{2u^3\over 3}-{u^2\over 6}-{3\over
u}-{1\over 2u^2}-{3\over 4u^3}\bigg)\,, \end{equation}

\begin{equation} {\rm Im}(\mu^2W_T-W_S)={g_A^2c_4^2m_\pi^6\over 8960\pi f_\pi^6}
(u-3)^3\bigg({1\over 2u^2}+{1\over 2u^3}-{u^3\over 18}-{u^2\over 2}-{5u\over 6}
-{4\over 3}-{11\over 6u}\bigg)\,. \end{equation}

\subsubsection{Spectral functions for $3\pi$-exchange class (b)}
\label{sec_3pib}

This class is displayed in Fig.~\ref{fig_dia3}(b). Each $3\pi$-exchange 
diagram of this class includes the one-loop $\pi N$-amplitude (completed by 
the low-energy constants $\bar d_j$). Only those parts of the 
$\pi N$-scattering amplitude, which are either independent of the pion 
cms-energy $\omega$ or depend on it linearly could be treated with the 
techniques available. The contributions are, in  general, small. Below, we 
present only the larger portions within this class. The omitted pieces are 
about one order of magnitude smaller. To facilitate a better understanding, 
we have subdivided this class into sub-classes labeled by roman numerals, 
following  Refs.~\cite{Kai01,Ent15}.

The auxiliary function
\begin{eqnarray}G(w) &=&\bigg[1+2g_A^2-{w^2\over 4}(1+5g_A^2)\bigg]{\sqrt{
w^2-4}\over w} \ln{w+\sqrt{w^2-4}\over 2}\nonumber \\ && +{w^2\over 24}
(5+13g_A^2)-1-2g_A^2+48\pi^2f_\pi^2\Big[(2-w^2)(\bar d_1+\bar d_2)+4\bar
d_5\Big]\,,\end{eqnarray}
arises from the part linear in $\omega$ of the isovector non-spin-flip 
$\pi N$-amplitude $g^-(\omega, t)$ with $t=(w m_\pi)^2$ (see e.g. Appendix B 
in Ref.~\cite{KGE12}). The spectral functions derived from this selected 
set of $3\pi$-exchange diagrams read as follows.

Class Xb:
\begin{equation} {\rm Im}W_S= {g_A^2m_\pi^6 \over (4f_\pi)^8\pi^5}\int_2^{u-1}
\!\!dw\,{4G(w)\over 27w^2u^4}\Big[(w^2-4)\lambda(w)\Big]^{3/2}\,,
\end{equation}

\begin{equation}{\rm Im}(\mu^2W_T-W_S)= {g_A^2m_\pi^6 \over (4f_\pi)^8\pi^5}
\int_2^{u-1}\!\!dw\, {4G(w)\over 9w^2u^4}(w^2-4)^{3/2} \sqrt{\lambda(w)}\,
{3u^2+1 \over u^2-1}\Big[u^4-(w^2-1)^2\Big]\,. \end{equation}

Class XIb:
\begin{equation} {\rm Im}W_S= {g_A^2m_\pi^6 \over (4f_\pi)^8\pi^5}\int_2^{u-1}
\!\!dw\,{8G(w)\over 27w^2u^4}(w^2-4)^{3/2} \sqrt{\lambda(w)}\, \Big[
2u^2(1+7w^2)-u^4-(w^2-1)^2\Big]\,, \end{equation}

\begin{equation}{\rm Im}(\mu^2W_T-W_S)= {g_A^2m_\pi^6 \over (4f_\pi)^8\pi^5}
\int_2^{u-1}\!\!dw\, {8G(w)\over 9w^2u^4}{(w^2-4)^{3/2}\over\sqrt{\lambda(w)}}
\,(u^2+1-w^2)^2\Big[2w^2(1+3u^2)-w^4-(u^2-1)^2\Big]\,.\end{equation}

Class XIIb:
\begin{equation} {\rm Im}W_S= {g_A^2m_\pi^6\over 9f_\pi^8(4\pi)^5}\int\!\!\!\!
\int\limits_{\!\!\!\!\!z^2<1}\!d\omega_1d\omega_2\,G(w)\Big[(\omega_1^2+\omega_2^2
-2)(1-3z^2)-5k_1 k_2 z \Big] \,, \end{equation}

\begin{equation} {\rm Im}(\mu^2W_T-W_S)= -{g_A^2m_\pi^6\over 3f_\pi^8(4\pi)^5}
\int\!\!\!\!\int\limits_{\!\!\!\!\!z^2<1}\!d\omega_1d\omega_2\,G(w)\omega_1
\omega_2\bigg[5+2z\bigg({k_1 \over k_2}+{k_2\over k_1}\bigg)\bigg] \,,
\end{equation}
setting $w = \sqrt{1+u^2-2u \omega_1}$.

Class XIIIb:
\begin{equation} {\rm Im}V_S= {g_A^4m_\pi^6 \over (4f_\pi)^8\pi^3u^3}\int_2^{u-1}
\!\!dw\,2G(w) \lambda(w)(2-w^2)\,, \end{equation}

\begin{equation} {\rm Im}(\mu^2V_T-V_S)= {g_A^4m_\pi^6 \over (4f_\pi)^8\pi^3u^3}
\int_2^{u-1}\!\!dw\,4G(w) (2-w^2) (1+u^2-w^2)^2\,, \end{equation}

\begin{eqnarray} {\rm Im}W_S&=& {g_A^4m_\pi^6\over 3f_\pi^8(4\pi)^5}\int\!\!\!\!
\int\limits_{\!\!\!\!\!z^2<1}\!d\omega_1d\omega_2\,G(w)\, \bigg\{u(\omega_1
+4\omega_2)-2-{\omega_1^2+8\omega_2^2\over 3}+z^2(\omega_1^2+4\omega_2^2-5)
\nonumber \\ &&+ {z k_2 \over k_1}(4u \omega_1+\omega_1^2-5)+{z k_1 \over k_2}
(u \omega_2+\omega_2^2-2)+{\arcsin(z) \over \sqrt{1-z^2}} \bigg[{k_1\over k_2}
(1-u \omega_2)+z(1-u \omega_1)\bigg]\bigg\}\,,\nonumber \\ \end{eqnarray}

\begin{eqnarray} && {\rm Im}(\mu^2 W_T-W_S)= {g_A^4m_\pi^6\over f_\pi^8(4\pi)^5}
\int\!\!\!\!\int\limits_{\!\!\!\!\!z^2<1}\!d\omega_1d\omega_2\,{2 \omega_1 \over 3}
G(w)\bigg\{ {2\omega_2 \over k_1^2}\Big[\omega_1(u-\omega_2)-1\Big]+u+
2\omega_2\nonumber \\ &&+{z k_1 \omega_2 \over k_2}+{z k_2 \over k_1}(4u+
\omega_1)+\omega_1\Big({2z k_2 \over k_1}\Big)^2+{\arcsin(z) \over k_1 k_2
\sqrt{1-z^2}}\bigg[(1+u^2)\Big(\omega_1+\omega_2-{u\over 2}\Big)
-2u \omega_1 \omega_2\bigg]\bigg\}\,, \end{eqnarray}
setting again $w = \sqrt{1+u^2-2u \omega_1}$.

Class XIVb:
\begin{equation} {\rm Im}V_S= {g_A^4m_\pi^6 \over (4f_\pi)^8\pi^3u^3}\int_2^{u-1}
\!\!dw\,{G(w)\over 2}\, \lambda(w)\Big[u^2+w^2+4(u^2-1)w^{-2}-5\Big]\,,
\end{equation}

\begin{equation} {\rm Im}(\mu^2V_T-V_S)= {g_A^4m_\pi^6 \over (4f_\pi)^8\pi^3u^3}
\int_2^{u-1}\!\!dw\,G(w)(w^2-1-u^2)\Big[w^4-2w^2(3+u^2)+(u^2-1)^2(1+4w^{-2})\Big]
\,. \end{equation}

\subsection{Four-pion exchange at N$^5$LO}
The exchange of four pions between two nucleons occurs for the first time at 
N$^5$LO. The pertinent diagrams involve three loops and only leading order 
vertices, which explains the sixth power in small momenta. Three-pion 
exchange with just leading order vertices turned out to be negligibly 
small~\cite{Kai00a,Kai00b}, and so we expect four-pion exchange with 
leading order vertices to be even smaller. Therefore, we can safely neglect 
this contribution.

\section{Perturbative $N\!N$-scattering in peripheral partial 
waves}\label{sec_pertNN}

\begin{table}
\caption{Low-energy constants as determined in Ref.~\cite{KGE12}.
The sets `GW' and `KH' are based upon the $\pi N$ partial wave analyses of 
Refs.~\cite{Arn06} and \cite{Koc86}, respectively. The $c_i$, $\bar{d}_i$, 
and $\bar{e}_i$ are in units of GeV$^{-1}$, GeV$^{-2}$, and GeV$^{-3}$.}
\begin{tabular}{crr}
\hline
\hline
           & \hspace*{5cm} GW & \hspace*{5cm} KH \\
\hline
$c_1$ & --1.13 & --0.75 \\
$c_2$ & 3.69 & 3.49 \\
$c_3$ & --5.51 & --4.77 \\
$c_4$ & 3.71 & 3.34 \\
$\bar{d}_1 + \bar{d}_2$ & 5.57 & 6.21 \\
$\bar{d}_3$ & --5.35 & --6.83 \\
$\bar{d}_5$ & 0.02 & 0.78 \\
$\bar{d}_{14} - \bar{d}_{15}$ & --10.26 & --12.02 \\
$\bar{e}_{14}$ & 1.75 & 1.52 \\
$\bar{e}_{15}$ & --5.80 & --10.41 \\
$\bar{e}_{16}$ & 1.76 & 6.08 \\
$\bar{e}_{17}$ & --0.58 & --0.37 \\
$\bar{e}_{18}$ & 0.96 & 3.26 \\
\hline
\hline
\end{tabular}
\label{tab_lecs}
\end{table}

\begin{figure*}
\vspace*{-2cm}
\scalebox{0.75}{\includegraphics{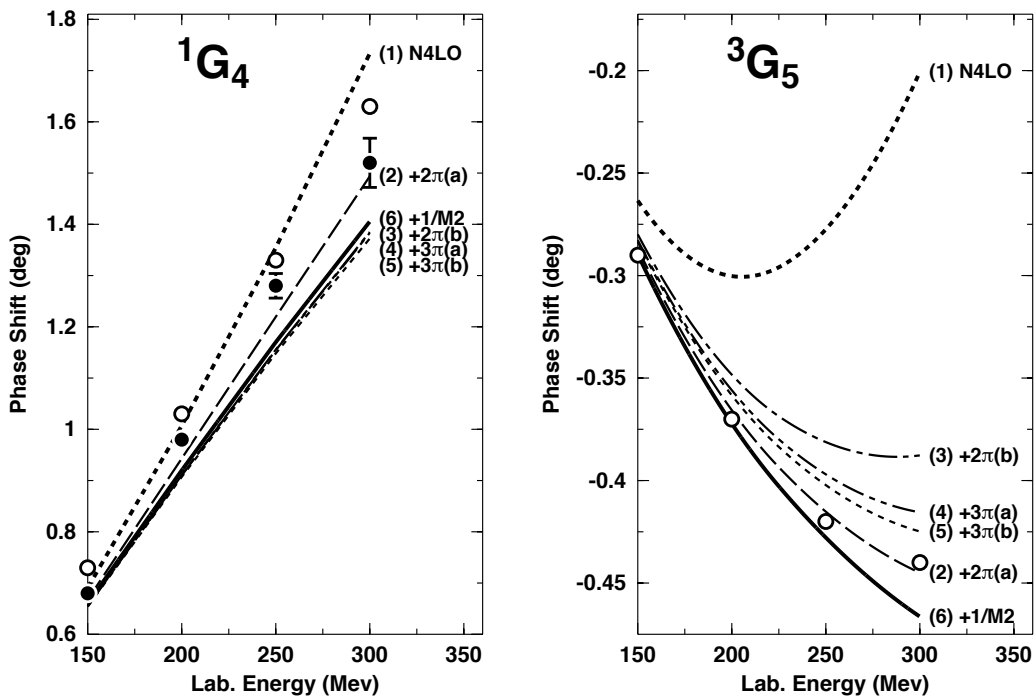}}
\vspace*{-9.0cm}
\caption{Effect of individual sixth-order contributions on the neutron-proton 
phase shifts of two important $G$-waves. The individual contributions are 
added up successively in the order given in parentheses next to each curve. 
Curve (1) is N$^4$LO and curve (6) contains all N$^5$LO contributions
calculated in this work. A SFR cutoff $\tilde{\Lambda}=900$ MeV is applied.
The filled and open circles represent the results from the Nijmegen 
multi-energy $np$ phase-shift analysis~\cite{Sto93} and the GWU $np$-analysis 
SP07~\cite{SP07}, respectively.\label{fig_ph1}}
\end{figure*}

To obtain an idea of the physical relevance and implications of the 
contributions evaluated in Sec.~II, we will now calculate the impact of these 
on elastic $N\!N$-scattering in peripheral partial waves. Specifically, we will 
consider partial waves with orbital angular momentum $L\geq 4$ (i.e., 
$G$-waves and higher), because polynomial terms up to sixth power do not make 
any contributions to these angular momentum states. The $L\geq 4$ partial 
waves are sensitive only to the non-polynomial pion-exchange expressions 
governed by chiral symmetry.

The smallness of the phase-shifts in peripheral partial waves suggests that 
the calculation can be done perturbatively. This avoids the complications
and possible model-dependences (e.g., cutoff-dependence) that the 
non-perturbative treatment with the Lippmann-Schwinger equation, necessary 
for low partial waves, is beset with.

Previous systematic investigations of peripheral partial waves have been 
conducted at N$^2$LO in Refs.~\cite{KBW97,KGW98}, at N$^3$LO in 
Ref.~\cite{EM02}, and at N$^4$LO in Ref.~\cite{Ent15}. Here, we will now 
present the investigation at N$^5$LO. 

The perturbative $K$-matrix for neutron-proton ($np$) scattering
is calculated as follows:
\begin{eqnarray}
K({\vec p}~',\vec p) &=& V_{1\pi}^{(np)}({\vec p}~',\vec p\,)+V_{2\pi, \rm it}^{(np)}
({\vec p}~',{\vec p}\,) + V({\vec p}~',{\vec p}\,) \label{eq_kmat}
\end{eqnarray}
with $V_{1\pi}^{(np)} ({\vec p}~', \vec p\,)$ the one-pion-exchange (1PE) 
potential that applies to $np$ scattering taking charge-dependence into 
account. It is given by
\begin{equation} V_{1\pi}^{(np)} ({\vec p}~', \vec p\,) 
= -V_{1\pi} (m_{\pi^0}) + (-1)^{I+1}\, 2\, V_{1\pi} (m_{\pi^\pm}) \,, \label{eq_1penp}
\end{equation}
where $I=0,1$ denotes the total isospin of the $pn$-system and
\begin{equation} V_{1\pi} (m_\pi) = - \frac{g_A^2}{4f_\pi^2} \, \frac{
\vec \sigma_1 \cdot \vec q \,\, \vec \sigma_2 \cdot \vec q} {q^2 + m_\pi^2} \,.
\end{equation}
We use the values $m_{\pi^0}=134.9766$ MeV and $m_{\pi^\pm}=139.5702$ MeV for 
the neutral and charged pion mass. $V_{2\pi, \rm it}^{(np)}({\vec p}~',\vec p\,)$ 
represents the once-iterated 1PE given by:
\begin{equation}
V_{2\pi, \rm it}^{(np)} ({\vec p}~',{\vec p}\,)  = {\cal P}\!\!\int {d^3p'' \over 
(2\pi)^3} \:\frac{M_N^2}{E_{p''}} \: \frac{V_{1\pi}^{(np)}({\vec p}~',{\vec p}~'')
\,V_{1\pi}^{(np)}({\vec p}~'',\vec p\,)} {{ p}^{2}-{p''}^{2}}\,, \label{eq_2piit}
\end{equation}
where ${\cal P}$ denotes the principal value and $E_{p''}=\sqrt{M_N^2+{p''}^2}$.
At sixth order, up to three iterations of $1\pi$-exchange should be included.
However, we found that the difference between the once-iterated 1PE and the 
infinitely-iterated 1PE is so small that it could not be identified on the 
scale of our phase shift figures. For that reason, we omit iterations of 1PE 
beyond what is contained in $V_{2\pi, \rm it}^{(np)} ({\vec p}~',\vec p\,)$.

Finally, the third term on the right hand side of Eq.~(\ref{eq_kmat}), 
$V({\vec p}~',\vec p\,) $, stands for the sum of irreducible multi-pion 
exchange contributions that occur at the order up to which the calculation 
is conducted. In multi-pion exchanges, we use the average pion mass $m_\pi = 
138.039$ MeV and, thus, neglect the charge-dependence due to pion-mass 
splitting. For the average nucleon mass, we use twice the reduced mass of 
the $pn$-system:
\beq M_N  =  \frac{2M_pM_n}{M_p+M_n} = 938.9183 \mbox{ MeV.} \eeq
Through relativistic kinematics, the CMS on-shell momentum $p$ is related to
the kinetic energy $T_{\rm lab}$ of the incident neutron in the laboratory 
system, by:
\beq
p^2  =  \frac{M_p^2 T_{\rm lab} (T_{\rm lab} + 2M_n)}
               {(M_p + M_n)^2 + 2T_{\rm lab} M_p}  \,,
\eeq
with $M_p=938.2720$ MeV and $M_n=939.5654$ MeV the proton and neutron masses, 
respectively. The $K$-matrix, Eq.~(\ref{eq_kmat}), is decomposed into partial 
waves following  Ref.~\cite{EAH71} and phase-shifts $\delta_L$ are then 
calculated via
\begin{equation}
\tan \delta_L (T_{\rm lab}) = -\frac{M_N^2p }{16\pi^2E_p} \, p \, K_L(p,p)\,.
\end{equation}
For more details concerning the evaluation of phase shifts, including the 
case of coupled partial waves, see Ref.~\cite{Mac93} or the appendix of 
Ref.~\cite{Mac01}.

Chiral symmetry establishes a link between the dynamics in the $\pi N$-system 
and the $N\!N$-system (through common low-energy constants). In order to check 
the consistency, we use the LECs for subleading $\pi N$-couplings as 
determined in analyses of low-energy elastic $\pi N$-scattering.
Appropriate analyses for our purposes are contained in 
Refs.~\cite{KGE12,WCE14}, where $\pi N$-scattering has been calculated at fourth 
order using the same power-counting of relativistic $1/M_N$-corrections as in 
the present work. Ref.~\cite{KGE12} performed two fits, one to the 
GW~\cite{Arn06} and one to the KH~\cite{Koc86} partial wave analysis 
resulting in the two sets of LECs listed in Table~\ref{tab_lecs}.
In our present work, we apply the GW set unless noted otherwise.
Moreover, we absorb the Goldberger-Treiman discrepancy into an effective 
value of the nucleon axial-vector coupling constant $g_A= g_{\pi NN}f_\pi/M_N
=1.29$.

\begin{figure*}
\vspace*{-1.5cm}
\hspace*{-0.7cm}
\scalebox{0.45}{\includegraphics{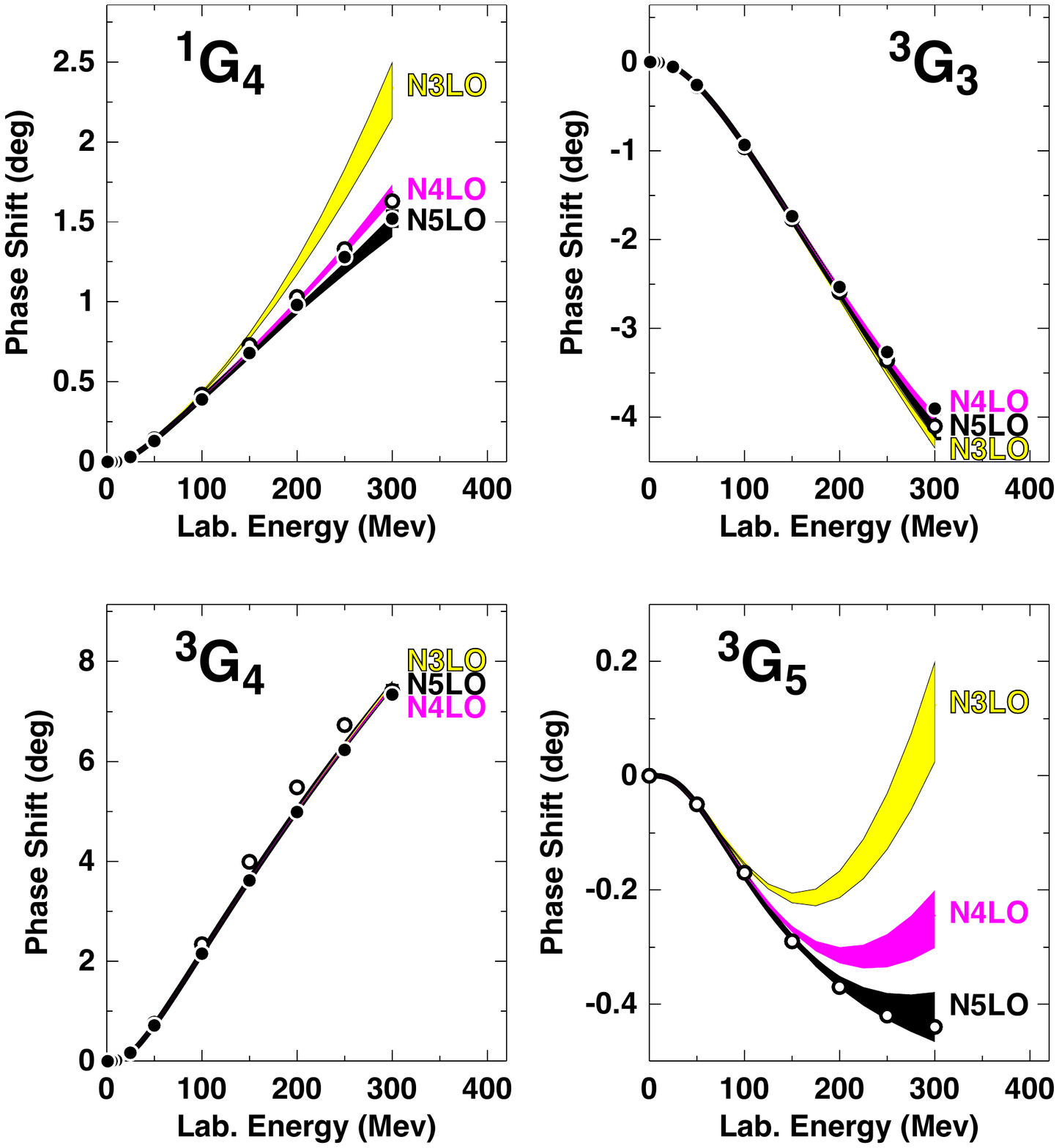}}
\hspace*{-1.3cm}
\scalebox{0.45}{\includegraphics{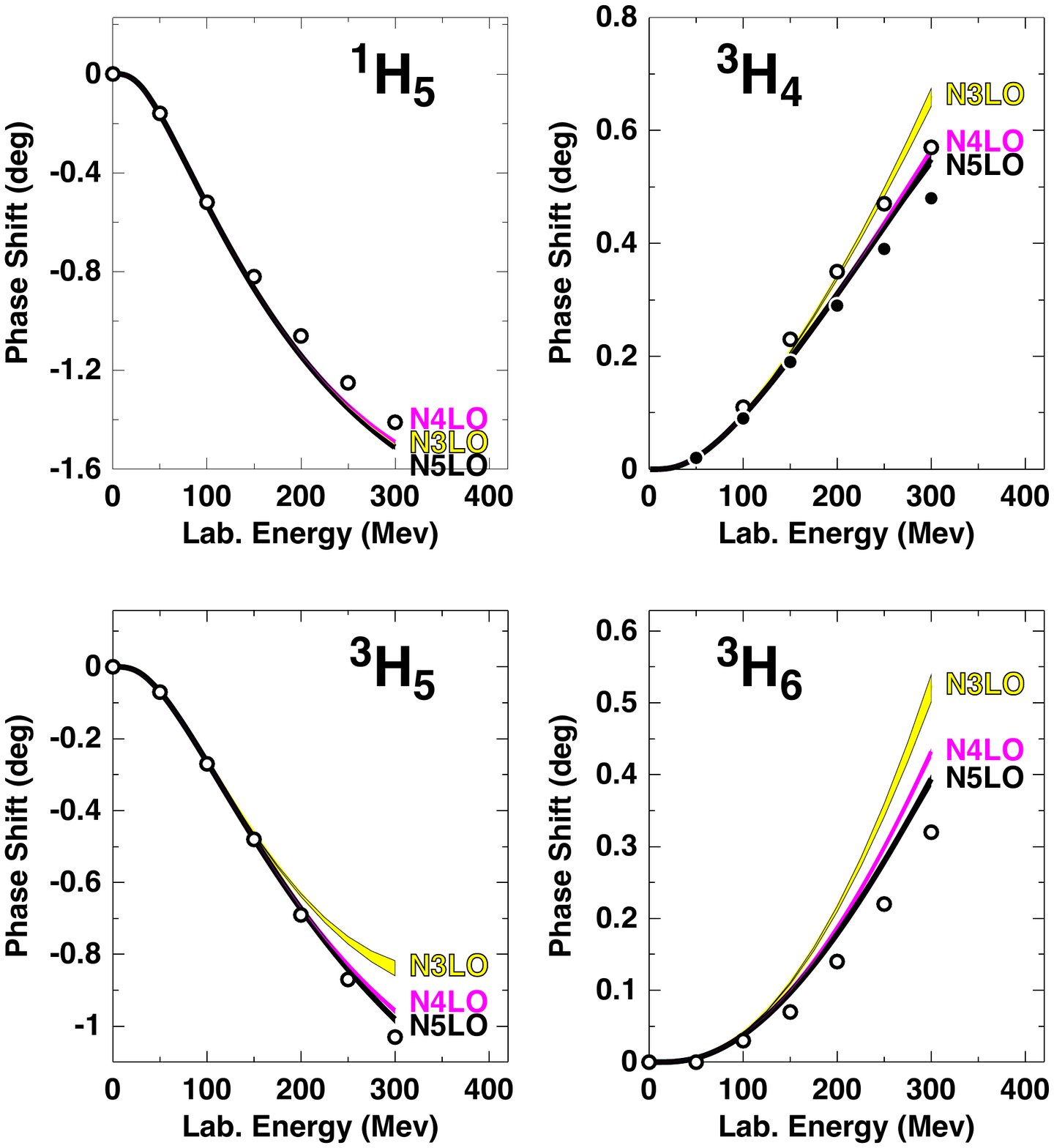}}
\vspace*{-1.5cm}
\caption{(Color online)
Phase-shifts of neutron-proton scattering in $G$ and $H$ waves at various 
orders as denoted. The shaded (colored) bands show the variations of the 
predictions when the SFR cutoff $\tilde{\Lambda}$ is changed over the range 
700 to 900 MeV. Empirical phase shifts are as in Fig.~\ref{fig_ph1}.
\label{fig_ph2}}
\end{figure*}

As shown in Figs.~\ref{fig_dia1} to \ref{fig_dia3} and derived in 
Sec.~\ref{sec_pions}, the sixth-order corrections consists of several 
contributions. We will now demonstrate how the individual sixth-order 
contributions impact $N\!N$-phase-shifts in peripheral waves.
For this purpose, we display in Fig.~\ref{fig_ph1}  phase-shifts
for two important peripheral partial waves, namely, $^1G_4$, and $^3G_5$.
In each frame, the following curves are shown:
\begin{description}
\item[(1)]
N$^4$LO (as defined in Ref.~\cite{Ent15}).
\item[(2)] The previous curve plus
the N$^5$LO $2\pi$-exchange contributions of class (a),
Fig.~\ref{fig_dia1}(a) and Sec.~\ref{sec_2pia}.
\item[(3)] The previous curve plus
the N$^5$LO $2\pi$-exchange contributions of class (b),
Fig.~\ref{fig_dia1}(b) and Sec.~\ref{sec_2pib}.
\item[(4)] The previous curve plus
the N$^5$LO $3\pi$-exchange contributions of class (a),
Fig.~\ref{fig_dia3}(a) and Sec.~\ref{sec_3pia}.
\item[(5)] The previous curve plus
the N$^5$LO $3\pi$-exchange contributions of class (b),
Fig.~\ref{fig_dia3}(b) and Sec.~\ref{sec_3pib}.
\item[(6)] The previous curve plus
the $1/M_N^2$-corrections (denoted by `1/M2'),
Fig.~\ref{fig_dia2} and Sec.~\ref{sec_rel}.
\end{description}
In summary, the various curves add up successively the individual N$^5$LO 
contributions in the order indicated by the curve labels. The last curve in 
this series, curve (6), includes all N$^5$LO contributions calculated in this 
paper. For all curves of this figure a SFR cutoff $\tilde{\Lambda}=900$ MeV 
[cf. Eq.~(\ref{eq_disp})] is employed.

From Fig.~\ref{fig_ph1}, we make the following observations. 
The two-loop $2\pi$-exchange class (a), Fig.~\ref{fig_dia1}(a), generates a 
strong repulsive central force through the spectral function 
Eq.~(\ref{eq_2pia1}), while the spin-spin and tensor forces provided by this 
class, Eq.~(\ref{eq_2pia2}), are negligible. The fact that this class 
produces a relatively large contribution is not unexpected, since it is 
proportional to $c_i^2$. The $2\pi$-exchange contribution class (b), 
Fig.~\ref{fig_dia1}(b), creates a moderately repulsive central force as seen 
by its effect on $^1G_4$ and a noticeable tensor force as the impact on 
$^3G_5$ demonstrates. The $3\pi$-exchange class (a), Fig.~\ref{fig_dia3}(a), 
is negligible in $^1G_4$, but noticeable in $^3G_5$ and, therefore, it should 
not be neglected. This contribution is proportional to $c_i^2$, which 
suggests a non-negligible size but it is  typically  smaller than the 
corresponding $2\pi$-exchange contribution class (a). The $3\pi$-exchange 
class (b) contribution, Fig.~\ref{fig_dia3}(b), turns out to be negligible 
[see the difference between curve (4) and (5) in Fig.~\ref{fig_ph1}].
This may not be unexpected since it is a three-loop contribution with only 
leading-order vertices. Finally the relativistic $1/M_N^2$-corrections to the 
leading $2\pi$-exchange, Fig.~\ref{fig_dia2}, have a small but non-negligible 
impact, particularly in $^3G_5$.

The predictions for all $G$ and $H$ waves, are displayed in Fig.~\ref{fig_ph2} 
in terms of shaded (colored) bands that are generated by varying the SFR 
cutoff $\tilde{\Lambda}$ [cf. Eq.~(\ref{eq_disp})] between 700 and 900 MeV.
The figure clearly reveals that, at N$^3$LO, the predictions are, in general, 
too attractive. As demonstrated in Ref.~\cite{Ent15}, the N$^4$LO 
contribution, essentially, compensates this attractive surplus. Now, let us 
turn to the new result at N$^5$LO: it shows a moderate repulsive contribution 
bringing the final prediction right onto the data (i.e. empirical 
phase-shifts). Moreover, the N$^5$LO contribution is, in general, 
substantially smaller than the one at N$^4$LO, thus, establishing a clear 
signature of convergence of the chiral expansion.

\begin{figure*}
\vspace*{-1.5cm}
\hspace*{-0.7cm}
\scalebox{0.45}{\includegraphics{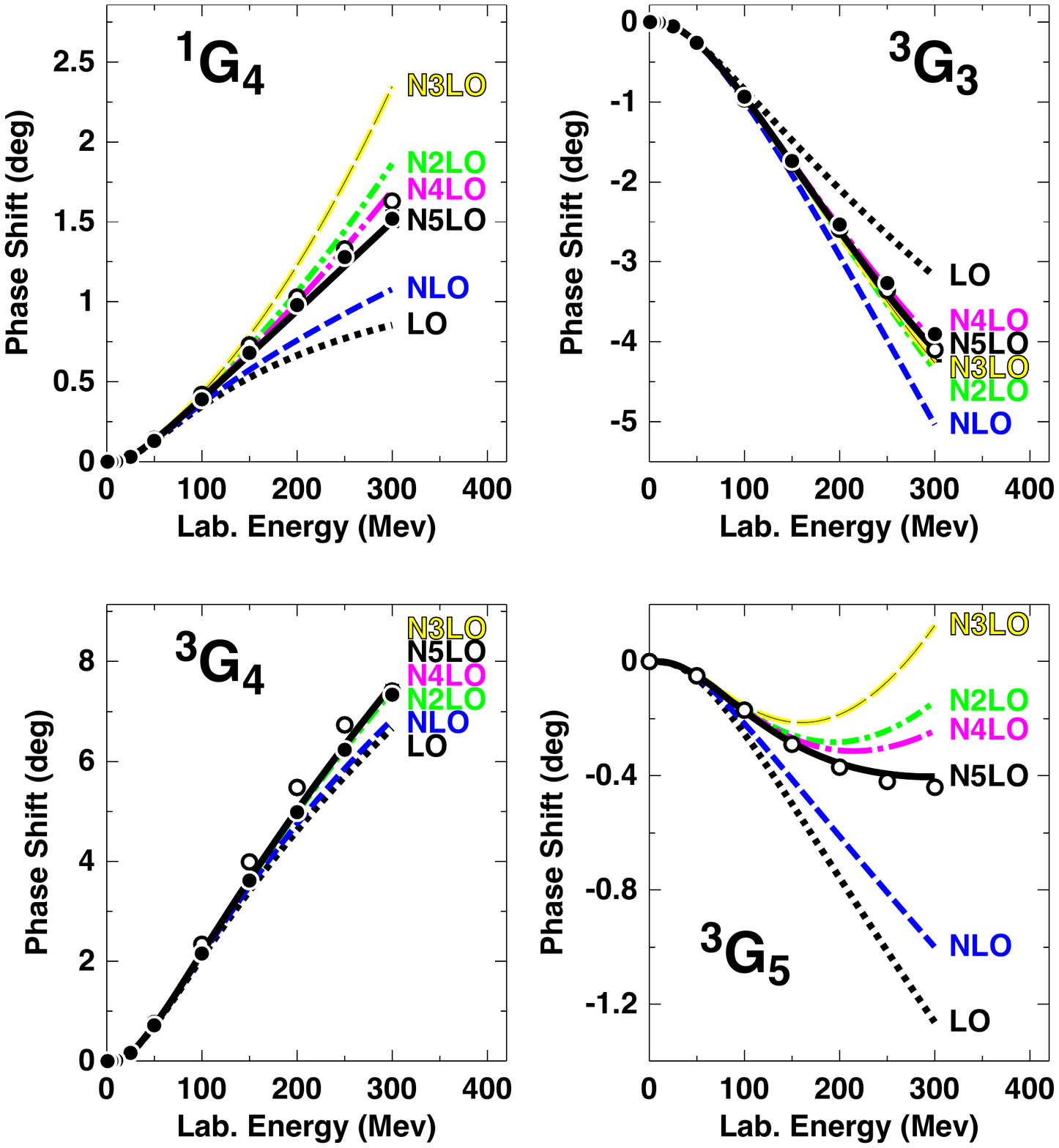}}
\hspace*{-1.2cm}
\scalebox{0.45}{\includegraphics{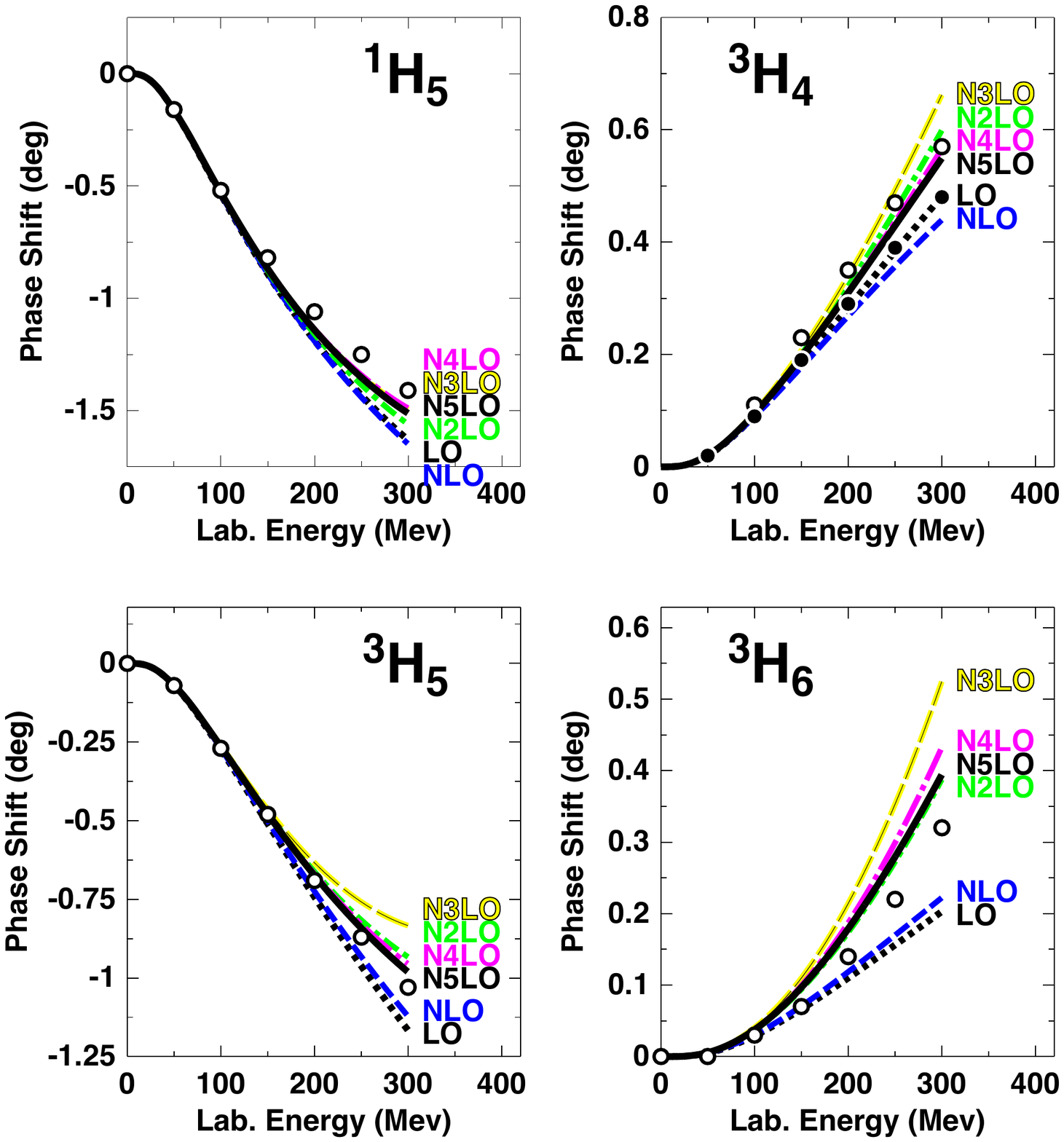}}
\vspace*{-1.5cm}
\caption{(Color online)
Phase-shifts of neutron-proton scattering in $G$ and $H$ waves at all orders 
from LO to N$^5$LO.  A SFR cutoff $\tilde{\Lambda}=800$ MeV is used.
Empirical phase shifts are as in Fig.~\ref{fig_ph1}.
\label{fig_ph3}}
\end{figure*}

\begin{figure*}
\vspace*{-1.5cm}
\scalebox{0.55}{\includegraphics{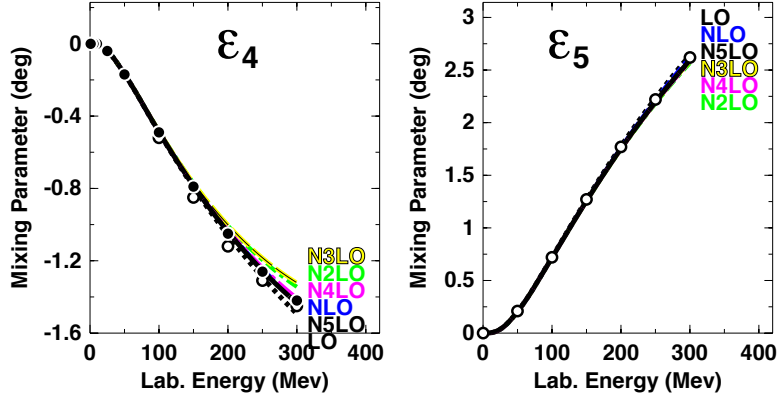}}
\vspace*{-8.0cm}
\caption{(Color online)
Mixing angles for neutron-proton scattering for $J=4, 5$ at all orders from 
LO to N$^5$LO.  A SFR cutoff $\tilde{\Lambda}=800$ MeV is used. Filled and 
open circles are as in Fig.~\ref{fig_ph1}.
\label{fig_ph4}}
\end{figure*}

\begin{figure*}
\vspace*{-1.5cm}
\scalebox{0.55}{\includegraphics{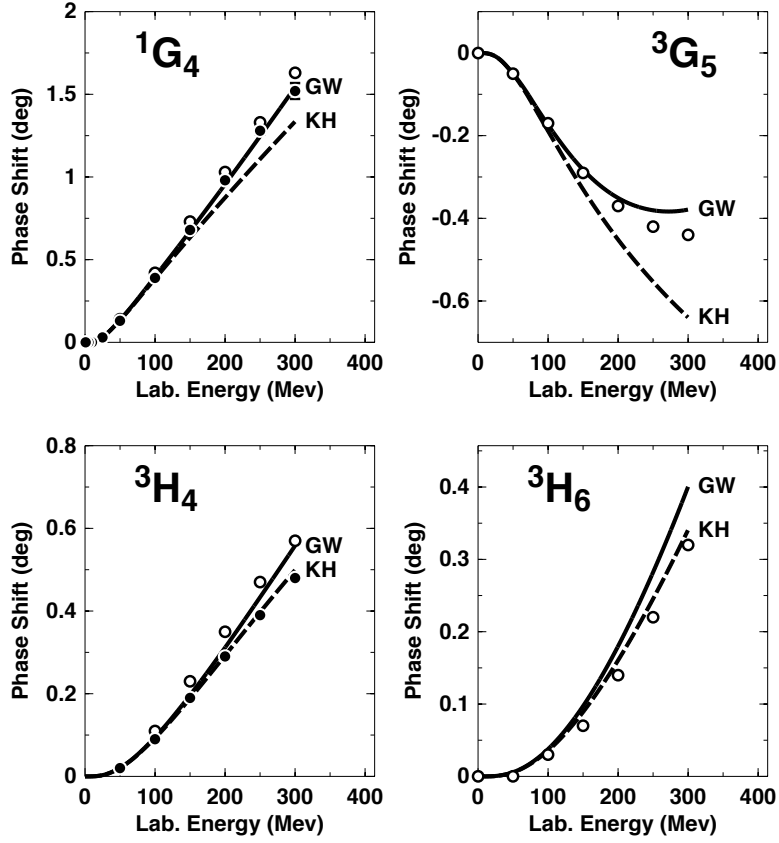}}
\vspace*{-2.0cm}
\caption{Neutron-proton phase shifts for some important $G$ and $H$ waves
using the GW and the KH sets of $\pi N$ LECs, as denoted. A SFR cutoff 
$\tilde{\Lambda}=700$ MeV is used. Empirical phase shifts are as in 
Fig.~\ref{fig_ph1}. \label{fig_ph5}}
\end{figure*}

At this point a comment is in place concerning the empirical phase shifts 
with which we compare our predictions in Figs.~\ref{fig_ph1} to \ref{fig_ph5}.
We use the 1993 Nijmegen analysis~\cite{Sto93} (represented by filled circles 
in the figures) and the GWU analysis from summer 2007~\cite{SP07} (open 
circles). We have also considered the recent Granada $N\!N$-analysis~\cite{PAA14}.
However, it turned out that, in general, the Granada and Nijmegen analyses 
are so close to each other that it does not make sense to show them 
separately. Concerning a second analysis, we decided for GWU~\cite{SP07} for 
two reasons. The GWU analysis is truly alternative to Nijmegen (and Granada), 
because it is not performed with a cleaned-up data base; it uses the full 
$N\!N$-data base. Moreover, the GWU analysis provides empirical phase shifts also 
for partial waves with $J=5, 6$, which we need. (The Nijmegen and Granada 
analyses stop at $J=4$.)

Figure~\ref{fig_ph2} includes only the three highest orders. However, a 
comparison between all orders is also of interest. Therefore, we show in 
Figs.~\ref{fig_ph3} the contributions to phase shifts  through all six
chiral orders from LO to N$^5$LO (as defined in Ref.~\cite{Ent15} and the 
present paper). Note that the difference between the LO prediction 
(one-pion-exchange, dotted line) and the data (filled and open circles) is to 
be provided by two- and three-pion exchanges, i.e. the intermediate-range part
of the nuclear force. How well that is accomplished is a crucial test for
any theory of nuclear forces.  NLO produces only a small contribution, 
but N$^2$LO creates substantial intermediate-range attraction (most clearly 
seen in $^1G_4$, $^3G_5$, and $^3H_6$). In fact, N$^2$LO is the largest 
contribution among all orders. This is due to the one-loop $2\pi$-exchange 
(2PE) triangle diagram which involves one $\pi\pi NN$-contact vertex 
proportional to $c_i$. This vertex represents correlated 2PE as well as 
intermediate $\Delta(1232)$-isobar excitation. It is well-known from the 
traditional meson theory of nuclear forces~\cite{MHE87,Vin79,Lac80}
that these two features are crucial for a realistic and quantitative 2PE model.
Consequently, the one-loop $2\pi$-exchange at N$^2$LO is attractive and 
assumes a realistic size describing the intermediate-range attraction of the
nuclear force about right. At N$^3$LO, more one-loop 2PE is added by the 
bubble diagram with two $c_i$-vertices, a contribution that seemingly is 
overestimating the attraction. This attractive surplus is then compensated by
the prevailingly repulsive two-loop $2\pi$- and $3\pi$-exchanges that occur 
at N$^4$LO and N$^5$LO. 

In this context, it is worth to note that also in conventional meson 
theory~\cite{MHE87} the one-loop models for the 2PE contribution always show 
some excess of attraction (cf. Figs.~7-9 of Ref.~\cite{EM02}). The same is 
true for the dispersion theoretic approach pursued by the Paris 
group~\cite{Vin79,Lac80}. In conventional meson theory, the surplus 
attraction is reduced by heavy-meson exchange ($\rho$- and $\omega$-exchange) 
which, however, has no place in chiral effective field theory (as a 
finite-range contribution). Instead, in the latter approach, two-loop 
$2\pi$- and $3\pi$-exchanges provide the corrective action.

We now turn to Figs.~\ref{fig_ph4}, where  we show how the six chiral orders
impact the mixing angles with $J=4, 5$. Note that the mixing angles 
depend only on the tensor force (the quadratic  spin-orbit term $V_{\sigma L}$ 
in Eq.(\ref{eq_4M2sL}) is very small). It is clearly seen that the 
$1\pi$-exchange (LO) alone describes these mixing angles correctly and that 
the various higher orders make only negligible contributions, particularly,
for $J=5$. At any order in the chiral expansion, tensor forces are created, 
but obviously the tensor force contributions beyond LO are of shorter range 
such that they do not matter in peripheral waves with $L\geq 4$.

In Figs.~\ref{fig_ph1} to \ref{fig_ph4} we employed the GW set of $\pi N$ 
LECs (cf.\ Table~\ref{tab_lecs}). Since these LECs carry some uncertainty~\cite{WCE14},
it is of interest to know what alternative sets will predict. In 
Fig.~\ref{fig_ph5} we show phase-shift predictions for the KH set and 
compare them to those from the GW set. It is seen that the differences are 
moderate and that both sets provide an about equally good description of the 
peripheral partial waves.

\section{Conclusions}
In this paper, we have calculated dominant $2\pi$- and $3\pi$-exchange 
contributions to the $N\!N$-interaction which occur at N$^5$LO (sixth order) of 
the chiral low-momentum expansion. The calculations are done in heavy-baryon 
chiral perturbation theory using the most general fourth order Lagrangian for 
pions and nucleons. We apply low-energy constants for subleading 
$\pi N$-coupling, which were determined from an analysis of elastic 
$\pi N$-scattering to fourth order using the same power counting scheme as 
in the present work. The spectral functions, which determine the 
$N\!N$-amplitudes via subtracted dispersion integrals, are regularized by a 
cutoff $\tilde{\Lambda}$ in the range 0.7 to 0.9 GeV. Besides the cutoff 
$\tilde{\Lambda}$, our calculations do not involve any adjustable parameters.

Recent work on $N\!N$-scattering in chiral perturbation theory~\cite{Ent15}, 
had revealed that the N$^2$LO, N$^3$LO, and N$^4$LO contributions are all 
about of the same size, thus raising some concern about the convergence of 
the chiral expansion for the $N\!N$-potential. Our present calculations show that 
the contribution at N$^5$LO is substantially smaller than the one at N$^4$LO, 
thus, establishing a clear signature of convergence. The two-loop 
$2\pi$-exchange contribution is the largest, while the corresponding 
three-loop contribution is small, but not negligible. Three-pion exchange is 
generally small at this order. The phase-shift predictions in $G$ and $H$ 
waves, where only the non-polynomial terms governed by chiral symmetry 
contribute, are in excellent agreement with the data.

This investigation represents the most comprehensive (and successful) test of 
the implications of chiral symmetry for the $N\!N$-system.

\section*{Acknowledgements}
This work was supported in part by the U.S. Department of Energy
under Grant No.~DE-FG02-03ER41270 (R.M. and Y.N.),
the Ministerio de Ciencia y
Tecnolog\'\i a under Contract No.~FPA2010-21750-C02-02 and
the European Community-Research Infrastructure Integrating
Activity ``Study of Strongly Interacting Matter'' (HadronPhysics3
Grant No.~283286) (D.R.E.), and by DFG and NSFC (CRC110) (N.K.).

\end{document}